\documentclass[runningheads]{llncs}

\usepackage{amsmath}
\usepackage{amsfonts}
\usepackage{mathtools}
\usepackage{multirow}
\usepackage{subfigure}
\usepackage{adjustbox}
\usepackage{wrapfig}
\usepackage{xspace}
\usepackage{algorithm}
\usepackage{algpseudocode}
\usepackage{csquotes}
\usepackage{booktabs}
\usepackage{xcolor}
\usepackage{enumerate}
\usepackage{bm}
\newcommand{\ie}{\emph{i.e.},\xspace}
\newcommand{\eg}{\emph{e.g.},\xspace}

\newcommand{\tb}[1]{\textbf{#1}}

\def\B#1{\mathbf #1}

\newcommand{\mn}{\texttt{DRAGON}\xspace}  
\begin{document}

\title{Enhancing Dyadic Relations with Homogeneous Graphs for Multimodal Recommendation}

\titlerunning{\mn}
%
\author{Hongyu Zhou \and
Xin Zhou \and
Lingzi Zhang \and
Zhiqi Shen}
%
%
\institute{Alibaba-NTU Singapore Joint Research Institute, \\ Nanyang Technological University, Singapore\\
\email{\{hongyu.zhou,xin.zhou,ZQShen\}@ntu.edu.sg, lingzi001@e.ntu.edu.sg}\\
}

\maketitle              
\begin{abstract}

User interaction data in recommender systems is a form of dyadic relation that reflects the preferences of users with items.
Learning the representations of these two discrete sets of objects, users and items, is critical for recommendation. 
Recent multimodal recommendation models leveraging multimodal features (\eg images and text descriptions) have been demonstrated to be effective in improving recommendation accuracy.
However, state-of-the-art models enhance the dyadic relations between users and items by considering either user-user or item-item relations, leaving the high-order relations of the other side (\ie users or items) unexplored.
Furthermore, we experimentally reveal that the current multimodality fusion methods in the state-of-the-art models may degrade their recommendation performance.
That is, without tainting the model architectures, these models can achieve even better recommendation accuracy with uni-modal information.
On top of the finding, we propose a model that enhances the dyadic relations by learning \underline{D}ual \underline{R}epresent\underline{A}tions of both users and items via constructing homogeneous \underline{G}raphs for multim\underline{O}dal recomme\underline{N}dation.
We name our model as \mn.
Specifically, \mn constructs the user-user graph based on the commonly interacted items and the item-item graph from item multimodal features.
It then utilizes graph learning on both the user-item heterogeneous graph and the homogeneous graphs (user-user and item-item) to obtain the dual representations of users and items.

To capture information from each modality,
\mn employs a simple yet effective fusion method, attentive concatenation, to derive the representations of users and items.
Extensive experiments on three public datasets and seven baselines show that \mn can outperform the strongest baseline by 22.03\% on average.
Various ablation studies are conducted on \mn to validate its effectiveness in modality fusing and learning dual representations for recommendation.
Our code is available at https://github.com/hongyurain/DRAGON.

\end{abstract}
\section{Introduction}

As society evolves, recommender systems have become indispensable tools to assist users to find products and services of their choice.
Previous work~\cite{adomavicius2005toward,rendle2012bpr,he2020lightgcn,zhang2022diffusion} explores historical user-item interactions that can be considered as a form of dyadic relation to capturing user preferences. 
However, these methods show inferior performance due to the sparse nature of interactions between users and items in real-world datasets.

To alleviate the data sparsity problem, recent multimodal recommender systems utilizing multimodal information (\eg item descriptive texts, product images) to improve recommendation performance have gained considerable attention.
A line of work~\cite{liu2017deepstyle,he2016vbpr} integrates multimodal features as side information to enhance latent item representations under the classic collaborative filtering framework. 
Inspired by the success of graph neural networks~(GNNs) on recommendation~\cite{wang2019neural,he2020lightgcn,zhou2022layer}, recent works focus on modeling user-item interactions as a bipartite graph and integrating multimodal information with graph structure. For example, MMGCN~\cite{wei2019mmgcn} builds a user-item bipartite graph for every modality to obtain modal-specific representation to understand user preference better. GRCN~\cite{wei2020graph} presents a graph refine layer that could identify noisy edges and corrupt false-positive edges to clarify the structure of the user-item interaction graph. DualGNN~\cite{wang2021dualgnn} and LATTICE~\cite{zhang2021mining} introduce either user-user or item-item relations into the user-item interactions and achieve state-of-the-art recommendation performance.
Although these models show effective recommendation accuracy, we argue the high-order relations in both sides of the dyadic relations can be explored simultaneously to fully address the data sparsity issue.
Inspired by the dual representation learning mechanism~\cite{zhang2022dual}, 
we enhance the representation learning of users and items by incorporating their dual representations to capture both the inter- and intra-relations between users and items. 

Furthermore, we experimentally reveal that these methods fail to fuse the modality features effectively. 
Specifically, we conduct an ablation study of multimodal features on two competitive multimodal models, DualGNN~\cite{wang2021dualgnn} and LATTICE~\cite{zhang2021mining}. 
The results shown in Table~\ref{tab: modality comparison} demonstrate that the performance of these models fed with a single modality, especially textual features, outperforms that with both modalities. 
This finding poses a meaningful question: \textit{How can we effectively fuse the multimodal information for recommendation?}

\begin{table}
\caption{Performance of DualGNN~\cite{wang2021dualgnn} and LATTICE~\cite{zhang2021mining} utilizing features in different modalities. R and N denote evaluation metrics Recall and NDCG. T and V denote textual and visual information.}
\small
\centering
\resizebox{0.8\columnwidth}{!}{
\begin{tabular}{l|c|ccc|ccc}
\hline
\multirow{2}*{Dataset} & \multirow{2}*{Metric} & \multicolumn{3}{c|}{DualGNN} & \multicolumn{3}{c}{LATTICE}\\
\cline{3-8}
~&~&V\&T&T&V&V\&T&T&V\\
\hline
\multirow{4}*{Baby} & R@10 & 0.0448 & \textbf{0.0612} & 0.0511 & \textbf{0.0547} & 0.0546 & 0.0492 \\
& R@20 & 0.0716 & \textbf{0.0943} & 0.0830 & 0.0850 & \textbf{0.0874} & 0.0781 \\
& N@10 & 0.0240 & \textbf{0.0331} & 0.0278 & \textbf{0.0292} & 0.0287 & 0.0265  \\
& N@20 & 0.0309 & \textbf{0.0417} & 0.0360 & 0.0370 & \textbf{0.0371} & 0.0339\\
\hline
\multirow{4}*{Sports} & R@10 & 0.0568 & \textbf{0.0697} & 0.0615 & 0.0620 & \textbf{0.0625} & 0.0572 \\
& R@20 & 0.0859 & \textbf{0.1060} & 0.0926 & 0.0953 & \textbf{0.0971} & 0.0887 \\
& N@10 & 0.0310 & \textbf{0.0379} & 0.0335 & 0.0335 & \textbf{0.0336} & 0.0312 \\
& N@20 & 0.0385 & \textbf{0.0473} & 0.0415 & 0.0421 & \textbf{0.0425} & 0.0393 \\
\hline
\multirow{4}*{Clothing} & R@10 & 0.0454 & \textbf{0.0524} & 0.0420 & 0.0492 & \textbf{0.0521} & 0.0408 \\
& R@20 & 0.0683 & \textbf{0.0798} & 0.0636 & 0.0733 & \textbf{0.0749} & 0.0614 \\
& N@10 & 0.0241 & \textbf{0.0281} & 0.0229 & 0.0268 & \textbf{0.0290} & 0.0221 \\
& N@20 & 0.0299 & \textbf{0.0351} & 0.0283  & 0.0330 & \textbf{0.0348} & 0.0273\\
\hline
\end{tabular}}
\label{tab: modality comparison}
\end{table}

To address the above question, we study the performance of different modality fusion methods, including \tb{Attentively Sum}, \tb{Max-pooling}, \tb{Mean-pooling}, and \tb{Concatenation}. Our experiments show that the late-fusion approach \tb{Concatenation} which directly concatenates the textual and visual features as the multimodal representation achieves the best performance.

Toward this end, we propose a framework that learns \underline{D}ual \underline{R}epresent\underline{A}tions of both users and items via constructing homogeneous \underline{G}raphs for multim\underline{O}dal recomme\underline{N}dation~(\mn). To be specific, \mn constructs a heterogeneous user-item bipartite graph for each modality to learn the modality-specific representations. It then utilizes the direct Concatenation fusion method to better exploit the learned modality-specific information. In order to learn the dual representations, we construct two homogeneous graphs based on the user co-occurrence and the item semantic features to capture the user preference from the neighbor users and the latent item content semantic from the neighbor items. Finally, \mn utilizes the learned dual representations of users and items to make recommendations. Extensive experiments are conducted on three public datasets to show the effectiveness of our proposed method. 

\section{Related Work}

\subsection{Multimodal Recommendation}
Collaborative filtering (CF) based models are widely used~\cite{he2017neural,wang2019neural,zhou2021selfcf,zhou2022layer} for recommender systems. The CF-based methods leverage the historical interactions between users and items to predict users' preferences. However, they usually suffer from the data sparsity problem since user-item interactions are generally limited for real-world datasets. 

To alleviate the data sparsity problem, massive multimodal content information has been utilized to improve recommendation performance. 
For example, VBPR~\cite{he2016vbpr} leverages the visual features from the pre-trained Convolutional Neural Network (CNN) to enhance the matrix factorization by incorporating the visual features with ID embeddings. 
VECF~\cite{chen2019personalized} models the user's attention on different regions of images and reviews.
Recently, GNNs have drawn more attention when applied to multimodal-based recommender systems. 
MMGCN~\cite{wei2019mmgcn} improves the quality of learned user and item representations by constructing a modality-specific user-item bipartite graph and adapting the message-passing mechanism of GNNs. 
Following MMGCN, GRCN~\cite{wei2020graph} presents a graph refine layer that can locate noisy edges and corrupt false-positive edges to refine the structure of the user-item interaction graph.
DualGNN\cite{zhang2022dual} adds an attention mechanism to capture the user's preference on different modalities. Meanwhile, it constructs a user-user graph to learn the user preference from neighbor users. 
LATTICE~\cite{zhang2021mining} builds an item-item graph for each modality, then combines them to create a latent modality-fused item graph. With graph convolution operations, items can share information from highly linked affinities in the graph to enhance their representations. 
Authors in~\cite{zhou2022tale} reveal that the learning of item-item graph shows negligible improvement on recommendation performance and freeze the graph for recommendation.
SLMRec~\cite{tao2022self} incorporates self-supervised learning tasks into the GNNs to uncover the latent patterns from multi-modalities to learn powerful representations. 
BM3~\cite{zhou2022bootstrap} bootstraps latent representations of both ID embeddings and multimodal features for effective and efficient recommendation.

\subsection{Multimodal Fusion}
Finding a fused multi-modal representation that is complimentary and comprehensible can greatly improve performance. Technically speaking, 
multi-modal fusion integrates information from different modalities to create a multimodal representation that can be applied to different tasks, such as link prediction~\cite{liu2021entity,liu2022m2gcn} and node classification~\cite{zhang2018multimodality}, etc. It can be categorized into early fusion, late fusion and hybrid fusion~\cite{baltruvsaitis2018multimodal}. Early fusion incorporates the extracted features at the beginning. Contrarily, the late fusion integrates after each modality has completed the decision (\eg classification or regression). Hybrid fusion combines the two methods.
For example, ACNet~\cite{hu2019acnet} adopts an early fusion method based on the attention mechanism. For the late fusion method, NMCL~\cite{wei2019neural} utilizes the cooperative networks for each modality to do feature augmentation with the attention mechanism, then a late fusion is applied to the prediction from various modalities. CELFT~\cite{wang2021combine} designs a hybrid fusion method that combines both early and late fusion together to overcome the shortcomings of the single fusion methods.

Multimodal recommendation models~\cite{wei2019mmgcn,zhang2021mining,zhang2022dual} normally apply mean-pooling, attentive sum or max-pooling. 
Our experimental results suggest that those models use a single modality representation leads to better performance than utilizing the multimodal representation learned from these fusion methods.
It indicates that pooling or attention-based methods may result in information loss when performing multimodal fusion. To mitigate this problem, we adopt attentive concatenation fusion but without reducing the embedding dimension, which has been demonstrated to be more effective in joining information from different modalities for recommendation.

\section{Methodology}
\begin{figure}
\centering
\includegraphics[width=\textwidth]{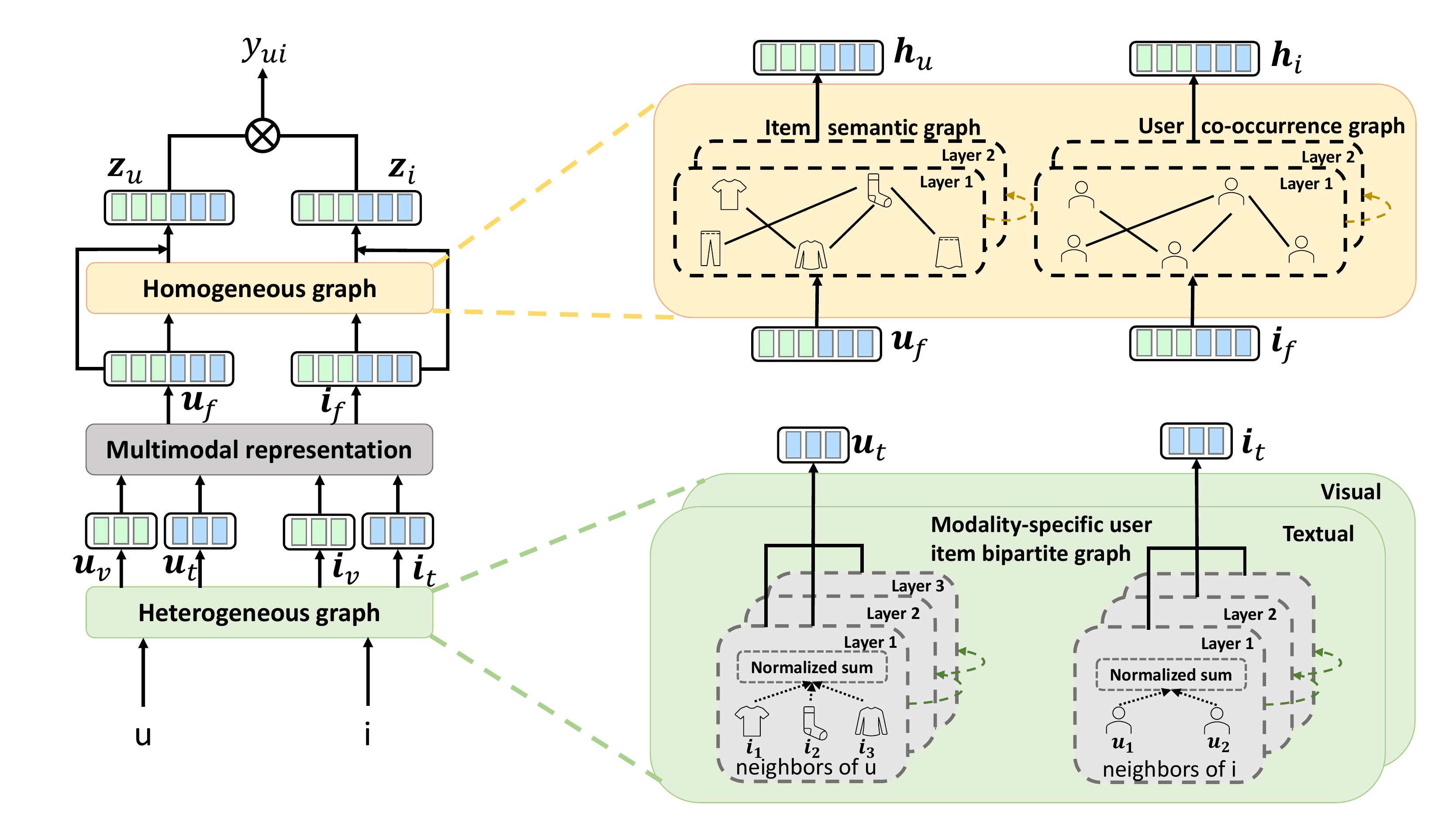}
\caption{The diagram depicts the main components of \mn. We first learn the modality-specific representations, then the multimodal representation learning module fused the single-modality representations and utilizes the homogeneous graph to capture the internal relations}
\label{structure}
\end{figure}

This section introduces the details of our proposed model \mn.
Fig.~\ref{structure} shows the overall architecture of \mn. There are four main components: 
(1) graph learning on a modality-specific heterogeneous graph to learn a uni-modal representation; 
(2) Multimodal representation learning module that captures user preference on different modalities and complementary information from each modality;
(3) graph learning on homogeneous graphs to capture the co-occurrence relation between users and the semantic relation between items; 
(4) a predictor module that ranks the candidate items based on the scores calculated from final user and item representations. 

\subsection{Preliminary}
Given a set of $N$ users $u \in \mathcal{U}$, a set of $M$ items $i \in \mathcal{I}$. 
We model the dyadic relations of user interactions as a user-item bipartite graph $\mathcal{G} = \{\mathcal{U},\mathcal{I}, \mathcal{E}\}$, 
where we regard the historical interactions as the set of edges in the graph denoted by $\mathcal{E}$. 
Besides the user-item interactions, each item is associated with multimodal content information $m \in \{v, t\}$, where $v$ and $t$ represent the visual and textual features respectively. 
We denote the modality feature for an item $i$ as $\bm{x}_{i}^{m}  \in \mathbb{R}^{d_{m} } $, where $d_{m}$ denotes the feature dimension of modality $m$.
In this paper, we only consider the visual and textual modalities denoted by $\mathit{v}$ and $\mathit{t}$. 
However, the proposed framework can be easily extended to scenarios involving more than two modalities.

\subsection{Dual Representation Learning}
Learning the representations of users and items is critical for the recommendation system. 
All representation learning-based techniques assume the existence of a common representation with consistent knowledge of different views of items~\cite{zhang2022dual}. Different views of an item contain specific discriminant information in addition to consistent knowledge about this item. We construct the heterogeneous and homogeneous graphs together to learn the dual representations of both user and item, which could capture both the internal association and the relationship between users and items.

\subsubsection{Heterogeneous Graph}
To learn modality-specific user and item representations, we construct a user-item graph for each modality, which is denoted as $\mathcal{G}_m$.
Following MMGCN~\cite{wei2019mmgcn}, we maintain the same graph structure $\mathcal{G}$ for different $\mathcal{G}_m$, but only keep the node features associated with a specific modality $m$.
We adopt LightGCN~\cite{he2020lightgcn} to encode $\mathcal{G}_m$. As shown in \cite{he2020lightgcn}, LightGCN simplifies the graph convolutional operations by excluding the feature transformation and nonlinear activation modules to improve recommendation performance and meanwhile ease the model optimization process.
Specifically, the user and item representations at the $(l+1)$-th graph convolution layer of $\mathcal{G}_m$ are derived as follows:
\begin{align}
        \bm{u}_{m}^{(l+1)} &= \sum_{i\in \mathcal{N}_{u} } ^{}\frac{1}{\sqrt{\left | \mathcal{N}_{u}  \right | }\sqrt{\left | \mathcal{N}_{i}  \right | } } \bm{i}_{m}^{(l)}, &
        \bm{i}_{m}^{(l+1)} &= \sum_{u\in \mathcal{N}_{i} } ^{}\frac{1}{\sqrt{\left | \mathcal{N}_{u}  \right | }\sqrt{\left | \mathcal{N}_{i}  \right | } } \bm{u}_{m}^{(l)},
\end{align}

Where $\mathcal{N}_u$ and $\mathcal{N}_i$ are the set of first hop neighbors of $u$ and $i$ in $\mathcal{G}_m$. $\bm{u}_{m}^{(0)}$ is randomly initialized and $\bm{i}_{m}^{(0)}$ is initialized with $\bm{x}_{i}^{m}$.
The symmetric normalization $\frac{1}{\sqrt{\left | N_{u}  \right | }\sqrt{\left | N_{i}  \right | } }$ is used to normalize the modality features learned from each layer which avoids the increase of scale when operates the graph convolutional operations.

After $L$ layers of data propagation, we combine the representations from every GCN layer using element-wise summation to derive the modality-specific representations for users and items. Formally,
\begin{align}
    \bm{u}_{m} &=\sum_{l=0}^{L} \bm{u}_{m}^{(l)}, & \bm{i}_{m} &=\sum_{l=0}^{L} \bm{i}_{m}^{(l)}.
    \label{eq2}
\end{align}

In such cases, the historical interactions and the modality information have been encoded into the final single-modal representations of users and items. And those operations have been applied to each modality by propagating on the modality-specific user-item bipartite graphs to learn the representations for each modality. 

\subsubsection{Homogeneous Graph}
In addition to employing the heterogeneous graph, which encodes the dyadic relation between users and items, we argue that the recommendation performance can be further enhanced by modeling the internal relations between users or items. For the two homogeneous graphs, we pre-established and freeze them to maintain the initial co-occurrence relation and semantic meaning.

\textit{User Co-occurrence Graph.} 
Based on the assumption that users who have interacted with similar items usually have similar preferences. We argue that the user's preference pattern is hidden inside the co-occurrence items and we construct a homogeneous user co-occurrence graph to learn the internal relations between users. However, the numbers of co-occurrence items between users are in a large range. In a general situation, the user will have a high number of commonly interacted items with a small group of users but with few items with other users. We only consider those who have more commonly interacted items with the user to capture similar preferences. 
To explicitly model the item co-occurrence patterns of users, we construct a homogeneous user co-occurrence graph $\tilde{\mathcal{G}}=\{\mathcal{U},\mathcal{P}_u\}$, where $\mathcal{P}_u=\{e_{u,u'}| u,u' \in \mathcal{U} \} $ denotes the edges between user nodes in $\tilde{\mathcal{G}}$ and $e_{u,u'}$ record the number of items that commonly interacted with u and u'. 

For every user $u\in \mathcal{U}$, we retain its top-$k$ users with the highest number of commonly interacted items. Specifically, we keep the edge weight $e_{u,u'}$ if $u'$ belongs to the top-$k$ users. Otherwise, the edge weight is 0.

\begin{equation}
        e_{u,u'} =
        \begin{cases}
        e_{u,u'}& \text{ if } e_{u,u'} \in \text{top-k}(e_{u}),\\
        0& \text{ otherwise.}
        \end{cases}
        \label{eq3}
\end{equation}

After establishing the user co-occurrence graph, we incorporate the attention mechanism when performing the graph propagation. The weight used for aggregating neighboring nodes for a user is computed using the softmax function to maximize the effect of neighboring users with a higher number of commonly interacted items. The representation of $u$ learned from $\tilde{\mathcal{G}}$ at layer $l+1$ is denoted as $\bm{h}_{u}^{(l+1)}$, which is derived as follows: 
\begin{align}
        \bm{h}_{u}^{(l+1)} =\sum_{u'\in \mathcal{N}_u} \frac{exp(e_{u,u'})}{\sum_{\hat{u}\in \mathcal{N}_u} exp(e_{u,\hat{u}})}{\bm{h}}_{u'}^{(l)},
    \label{eq4}
\end{align}

Where $e_{u,u'}$ indicates the number of common interacted items between $u$ and $u'$ and $\mathcal{N}_u$ denotes the neighbors of user $u$ in $\tilde{\mathcal{G}}$.
In this case, the representation of each user can be enhanced based on neighbors in the co-occurrence graph.

\textit{Item Semantic Graph.} 
Multimodal features offer rich and valuable content information about items, but previous studies~\cite{wei2020graph,wang2021dualgnn} neglect the significant underlying semantic relations of item features. 
Inspired by \cite{zhang2021mining}, we argue that item features are objective and we could establish the modality-specific homogeneous item graphs based on the raw features to learn the internal relations between items. Specifically, we construct the modality-aware item semantic graph $\hat{\mathcal{G}}_m=\{\mathcal{I},\mathcal{P}_m^i \}$ for each modality $m$, where $\mathcal{P}_m^i=\{e^m_{i,i'}| i,i' \in \mathcal{I} \} $ denotes the edges between item nodes in $\hat{\mathcal{G}}_m$. For an edge $e_{i,i'}$, its weight is calculated by the cosine similarity between original modality features of $\bm{x}_i^m$ and $\bm{x}_{i'}^m$:
\begin{align}
        e_{i,i'}^{m} =\frac{({\bm{x}_i^m})^{\top}{\bm{x}_{i'}^m}}{\| \bm{x}_i^m \| \| \bm{x}_{i'}^m\|  }.
        \label{eq5}
\end{align}

The derived $\hat{\mathcal{G}}_m$ is a fully connected graph where edge weights are calculated based on the similarity scores of modality features of connected nodes. Next, we make the graph sparse by retaining the top-$k$ similar items of every item. As the $\hat{\mathcal{G}}_m$ is a weighted graph, we convert it into an unweighted graph that captures the fundamental relation structure of the most related items~\cite{chen2009fast}. Formally,
\begin{equation}
    \begin{split}
        e_{i,i'}^{m} =
        \begin{cases}
        1& \text{ if } e_{i,i'}^{m} \in \text{top-k}(e_{i}^{m}),\\
        0& \text{ otherwise.}
        \end{cases}
    \end{split}
    \label{eq6}
\end{equation}

Since we get one item semantic graph for each modality, we combine them by performing weighted summation based on the importance score $\alpha _{m}$ that indicates the contribution of each modality and the summation is 1. Formally, $\hat{\mathcal{G}} = \{\mathcal{I}, \mathcal{P}_i\}$, where $\mathcal{P}_i=\{e_{i,i'}|i,i'\in\mathcal{I}\}$ and $e_{i,i'}=\sum_{m\in M}\alpha e^{m}_{i,i'}$.

After establishing the item semantic graph, we apply the graph convolution operation on it to capture the item-item relationship:
\begin{equation}
        \bm{h}_{i}^{(l+1)}=\sum_{i'\in \mathcal{N}_i} e_{i,i'}\bm{h}_{i'}^{(l)},
        \label{eq7}
\end{equation}

Where $\mathcal{N}_i$ denotes the neighbors of item $i$ in $\mathcal{G}^i$. Both $\bm{h}_{u}^{(0)}$ and $\bm{h}_{i}^{(0)}$ are initialized with their fused representations $\bm{u}_f$ and $\bm{i}_f$, which are introduced in the following section.

\subsection{Multimodal Fusion}
An important factor influencing multimodal recommendation accuracy is multimodal fusion. 
As we mentioned in Section 1, previous multimodal recommendation models utilizing single-modal information perform better than multimodal information.
We speculate their fusion methods may fail to capture modality-specific characteristics and even corrupt the learned single-modality representation.
We intend to learn the multi-modality that can capture complementary information which can not be contained in the single modality.
Specifically, to fuse the single modal features derived from modality-specific user-item graphs, we apply the \textbf{Attentive Concatenation} for user multimodal embeddings that capture the user preference on different modalities and direct concatenation for item multimodal embeddings. The attention weight $\alpha$ for users is initialized to 0.5. Formally,
\begin{align}
   \bm{u}_{f} &= \alpha \bm{u}_v\parallel (1-\alpha) \bm{u}_t, &
   \bm{i}_{f} &= \bm{i}_v\parallel \bm{i}_t,
   \label{eq8}
\end{align}

Where $\parallel$ denotes the concatenation operation. 
By performing attentive concatenation, we assume the single modality representations carry the richest information for each modality and this operation can capture the intact complementary information from each modality.

\subsection{Integration with Dual Representations}
We integrate representations of users and items learned from heterogeneous~(\ie Modality-specific User-Item Graphs) and homogeneous~(\ie User Co-occurrence Graph \& Item Semantic Graph) graphs to form their dual representations such that interactions between users and items and their internal relations can be well captured. 
We perform element-wise summation on the outputs learned from the three graphs to generate the dual representations for $u$ and $i$:
\begin{align}
        \bm{z}_u &= \bm{u}_{f} + \bm{h}_{u}^{L_{u}}, &
        \bm{z}_i &= \bm{i}_{f} + \bm{h}_{i}^{L_{i}},
\end{align}

Where $\bm{z}_u$ and $\bm{z}_i$ denote the final representations of user $u$ and item $i$. $L_{u}$ and $L_{i}$ denote the number of GCN layers for the user co-occurrence graph and item semantic graph respectively.

\subsection{Optimization} 
To optimize the parameters of \mn for the recommendation task, we leverage the Bayesian Personalized Ranking~(BPR) loss~\cite{rendle2012bpr}, which aims to score higher for the positive item than the negative one. We construct a triplet set $\mathcal{R}$ including the triplet $(u,i,j)$ for each user $u$ with the positive item $i$ and a randomly sampled negative item $j$ that has no interactions with $u$. The loss function $\mathcal{L}_{rec}$ is defined as follows,
\begin{align}
    \begin{split}
    &\mathcal{R}=\{(u,i,j)|(u,i) \in \mathcal{E},(u,j) \notin \mathcal{E}\},  \\
     &\mathcal{L}_{rec}=\sum_{(u,i,j) \in \mathcal{R}} -\textrm{ln}\sigma (y_{ui}-y_{uj}) +\lambda \| \B{\Theta} \|_2, 
    \end{split}
\end{align}

Where $y_{ui}=\bm{z}_u^{\top} \bm{z}_i$ calculates the inner product of $\bm{z}_u$ and $\bm{z}_i$, $\lambda$ is the $L_2$ regularization weight, and $\B{\Theta}$ denotes model parameters.

\section{Experiments}
\subsection{Experimental Settings}
\subsubsection{Datasets}
We conduct experiments on three categories Baby, Sports and Clothing from the Amazon dataset~\cite{mcauley2015image}, which contains product descriptions and images as textual and visual features. 
We retain the 5-core users and items such that each user or item is associated with at least 5 interactions, which is widely used in previous studies~\cite{he2016vbpr,he2020lightgcn,zhang2021mining}.
We use the pre-trained sentence-transformers~\cite{reimers2019sentence} to extract text features with a dimension equal to 384 and follow \cite{zhang2021mining} to use the published 4096-dimensional visual features. The statistics of datasets are summarised in Table~\ref{tab: dataset}. In this case, data sparsity~\cite{zhou2016evaluating} is calculated by dividing the number of interactions by the product of the number of items and users. 
\begin{table}
\caption{Statistics of the datasets.}
\small
\centering
\begin{tabular}{ l |r r r |r } 
\hline
Dataset & \# Users & \# Items & \# Interactions & Sparsity \\
\hline
Baby & 19,445 & 7,050 & 160,792 & 99.88\% \\
Sports & 35,598 & 18,357 & 296,337 & 99.95\% \\
Clothing & 39,387 & 23,033 & 278,677 & 99.97\%\\
\hline
\end{tabular}
\label{tab: dataset}
\end{table}
\subsubsection{Baselines}
To evaluate the performance of our proposed model \mn, we compare it with the following baseline methods which can be divided into two groups. 
\\
1) General recommendation models:
\begin{itemize}
    \item \textbf{BPR} \cite{rendle2012bpr} optimizes the user and item representations utilizing the matrix factorization method.
    \item \textbf{LightGCN} \cite{he2020lightgcn} simplifies the Graph Convolution Network by discarding the feature transformation and nonlinear activation modules.
\end{itemize} 
2) Multimodal recommendation models:
\begin{itemize}
    \item \textbf{VBPR} \cite{he2016vbpr} integrates visual features into item representations. For fair comparsion, we combine text and vision features to learn item representations.
    \item \textbf{DualGNN} \cite{wang2021dualgnn} proposes to use representations learned from modality-specific graphs and fuses the representations of neighbors in the user correlation graph.
    \item \textbf{GRCN} \cite{wei2020graph} locates and removes the false-positive edges in the graph. It then learns representations of items and users by conducting information propagation and aggregation in the refined graph.
    \item \textbf{LATTICE} \cite{zhang2021mining} introduces an item-item graph on each modality and obtains the latent item semantic graph by aggregating information from all modalities.
    \item \textbf{SLMRec} \cite{tao2022self} uses self-supervised learning techniques that supplements the supervised tasks to uncover the hidden signals from the data itself with contrastive loss.
\end{itemize} 

\subsubsection{Evaluation Metrics}
We follow the settings as previous models ~\cite{zhang2021mining,zhou2022layer,zhou2022tale} to randomly split the historical interactions with the ratio of 8:1:1 as train, valid and test sets. Moreover, we adopt the widely used metrics Recall@$K$ and NDCG@$K$ (denoted by R@$K$ and N@$K$) to evaluate the top-$K$ recommendation performance. 
We empirically set $K= 10$ and 20. 
For each metric, we compute the performance of each user in the testing data and report the average performance over all users. 

\subsubsection{Implementation Details}
We implement our proposed model by PyTorch~\cite{paszke2019pytorch} and embed the users and items with a dimensional size of 64 for all models. We use the Xavier method~\cite{glorot2010understanding} to initialize the embedding parameters, utilize Adam~\cite{kingma2014adam} as the optimizer, and fix the mini-batch size to 2048. All models are evaluated on a Tesla V100 32GB GPU card. The optimal hyper-parameters are determined via grid searches on the validation set: we do a grid search on the learning rate in \{1e-1, 1e-2, 1e-3, 1e-4, 1e-5,1e-6\} and the regularization weight in \{1e-1, 1e-2, 1e-3, 1e-4, 1e-5\}. We fix the number of GCN layers in the heterogeneous graph and homogeneous graph with ${L}=2$ and $L_{u}=L_{i}=1$, respectively. The important score $\alpha_m$ for image weight in the item semantic graph is empirically fixed as 0.1. The $k$ of top-$k$ in the user co-occurrence graph is set to 10. 
We set the maximum number of epochs to 1000 and adopt the early stopping strategy. That is, the model terminates when R@20 on the validation set does not increase for 20 successive epochs.
The model implementation has been integrated into the multimodal recommendation platform, MMRec~\cite{zhou2023mmrecsm}.

\begin{table}
\caption{Performance of baselines in terms of Recall and NDCG. Best results are in \textbf{boldface} and the second best is \underline{underlined}. ``\%Imp" denotes the relative improvement of \mn over the best baseline.}
\scriptsize
\centering
\def\arraystretch{1.0}
\begin{tabular}{l|c|cc|cccccc|c}
\hline
\multirow{2}*{Dataset} & \multirow{2}*{Metric} & \multicolumn{2}{c|}{General Model} & \multicolumn{6}{c|}{Multimodal Model} \\
\cline{3-10}
~&~&BPR&LightGCN&VBPR&DualGNN&GRCN&SLMRec&LATTICE&\mn& \%Imp\\


\hline
\multirow{4}*{Baby} & R@10 & 0.0357 & 0.0479 & 0.0423 & 0.0448 & 0.0539 & 0.0529 & \underline{0.0547} & \textbf{0.0662} & 21.02\%\\
& R@20 & 0.0575 & 0.0754 & 0.0663 & 0.0716 & 0.0833 & 0.0775 & \underline{0.0850} & \textbf{0.1021} & 20.12\%\\
& N@10 & 0.0192 & 0.0257 & 0.0223 & 0.0240 & 0.0288 & 0.0290 & \underline{0.0292} & \textbf{0.0345}& 18.15\%\\
& N@20 & 0.0249 & 0.0328 & 0.0284 & 0.0309 & 0.0363 & 0.0353 & \underline{0.0370}& \textbf{0.0435}&17.57\%\\
\hline
\multirow{4}*{Sports} & R@10 & 0.0432 & 0.0569 & 0.0558 & 0.0568 & 0.0598 & \underline{0.0663} & 0.0620&\textbf{0.0749} &12.97\% \\
& R@20 & 0.0653 & 0.0864 & 0.0856 & 0.0859 & 0.0915 & \underline{0.0990} & 0.0953&\textbf{0.1124} &13.54\%\\
& N@10 & 0.0241 & 0.0311 & 0.0307 & 0.0310 & 0.0332 & \underline{0.0365} &0.0335 &\textbf{0.0403} &10.41\%\\
& N@20 & 0.0298 & 0.0387 & 0.0384 & 0.0385 & 0.0414 & \underline{0.0450} & 0.0421&\textbf{0.0500} &11.11\%\\
\hline
\multirow{4}*{Clothing} & R@10 & 0.0206 & 0.0361 & 0.0281 & 0.0454 & 0.0424 & 0.0442 & \underline{0.0492} & \textbf{0.0650}&32.11\%\\
& R@20 & 0.0303 & 0.0544 & 0.0415 & 0.0683 & 0.0662 & 0.0659 & \underline{0.0733}& \textbf{0.0957}&30.56\%\\
& N@10 & 0.0114 & 0.0197 & 0.0158 & 0.0241 & 0.0223 & 0.0241 & \underline{0.0268}&\textbf{0.0357} &33.21\%\\
& N@20 & 0.0138 & 0.0243 & 0.0192  & 0.0299 & 0.0283 & 0.0296 & \underline{0.0330}& \textbf{0.0435}&31.82\%\\
\hline
\end{tabular}
\label{tab: performance comparison}
\end{table}

\subsection{Performance Comparison}
As shown in Table~\ref{tab: performance comparison}, we compare the recommendation performance of the state-of-the-art methods with our proposed model. 
We have the following observations:
\begin{itemize}
    \item All GCN-based methods perform better than MF-based recommendation models~(\ie BPR and VBPR), which demonstrates the effectiveness of modeling the historical interactions by a graph with graph convolutional operations. 
    \item In all evaluation metrics including Recall and NDCG, \mn outperforms all the baseline models across all the datasets. 
    For example, \mn makes an improvement over the strongest baseline in the term of R@10 on the datasets Baby, Sports and Clothing by 21.02\%, 12.97\% and 32.11\% respectively. 
    The improvement attributes to the dual representations and the multimodal fusion method. Learning dual representations from the heterogeneous and homogeneous graphs captures both the historical interactions and the internal relations among each set of dyadic objects~(\ie users or items). 
    A homogeneous graph helps to learn relevant characteristics from the neighbors. The fusion method helps to learn the complementary information from each single modal to enhance the multimodal representation.
    \item The multimodal recommendation models outperform the general recommendation models. GRCN, LATTICE and SLMRec are multimodal models which outperform all general methods. The performance of VBPR that builds upon the BPR framework by introducing modality information outperforms BPR on all datasets. However, there exist multimodal models that have a strong reliance on the representativeness of multimodal characteristics of items, which results in inconsistent performance across various datasets. 
    For example, DualGNN builds based on LightGCN and outperforms it on the Clothing dataset, but is less effective on Baby and Sports. It is possible that multimodal features are more important to reveal item characteristics in the clothing dataset, but are less informative in the other two datasets, in which DualGNN is inferior to LightGCN. 
\end{itemize}
In addition, we evaluate the scalability of \mn on a larger dataset Electronic of Amazon dataset with around 1.7M interactions, 200K users and 63K items. LATTICE~\cite{zhang2021mining} consumes more memory than the other baselines which could not be handled by the 32GB GPU card. SLMRec~\cite{tao2022self} needs to find the optimal parameter by grid search on more than 200 parameter sets which takes a long training time. \mn is superior to these baselines on the large graph. Although SLMRec is the strongest baseline on Electronic, \mn is capable of gaining an improvement of 4.76\% in terms of R@20 compared with SLMRec.

\subsection{Ablation Study}
In this section, we conduct exhaustive experiments to examine the behaviors of our proposed model under various settings.
\subsubsection{Effect of different components of \mn}
We design the following variants of \mn based on the homogeneous graphs used and compared with the strongest baselines (LATTICE and SLMRec) to investigate the contribution of different components of \mn:
\begin{itemize}
    \item $\mn_{UI}$ excludes the homogeneous graphs and only utilizes the heterogeneous graph;
    \item $\mn_{UU}$ incrementally adds the user co-occurrence graph to $\mn_{UI}$. This variant can capture the relations between users, which means only users have dual representations.
\end{itemize}
\begin{figure}
\centering
\subfigure[Comparison of different components]{
\includegraphics[width=3.8cm]{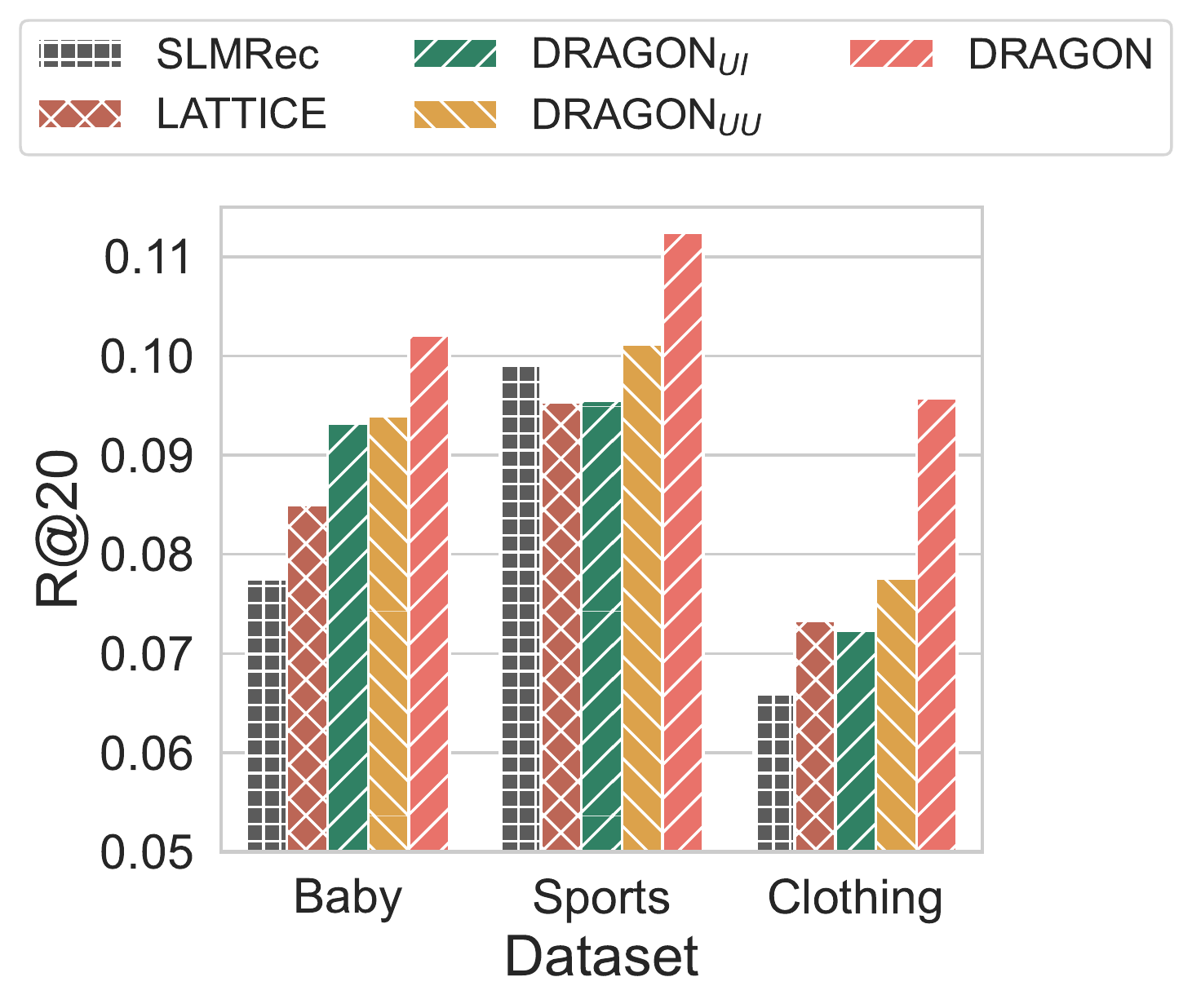}
\label{components}
}
\quad
\subfigure[Comparison of different fusion methods]{
\includegraphics[width=3.5cm]{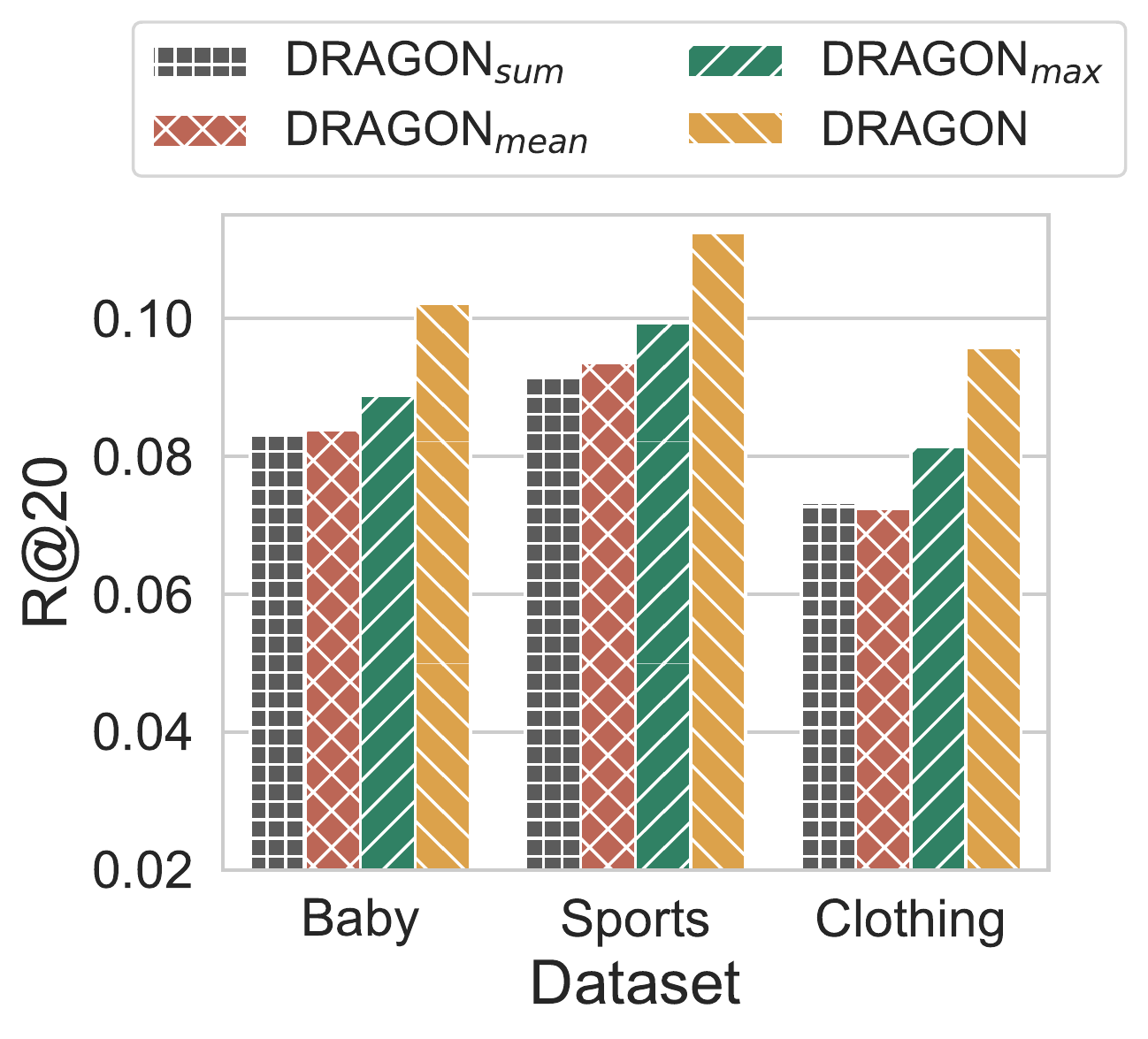}
\label{fusion}
}
\quad
\subfigure[Comparison of different modality]{
\includegraphics[width=3.4cm]{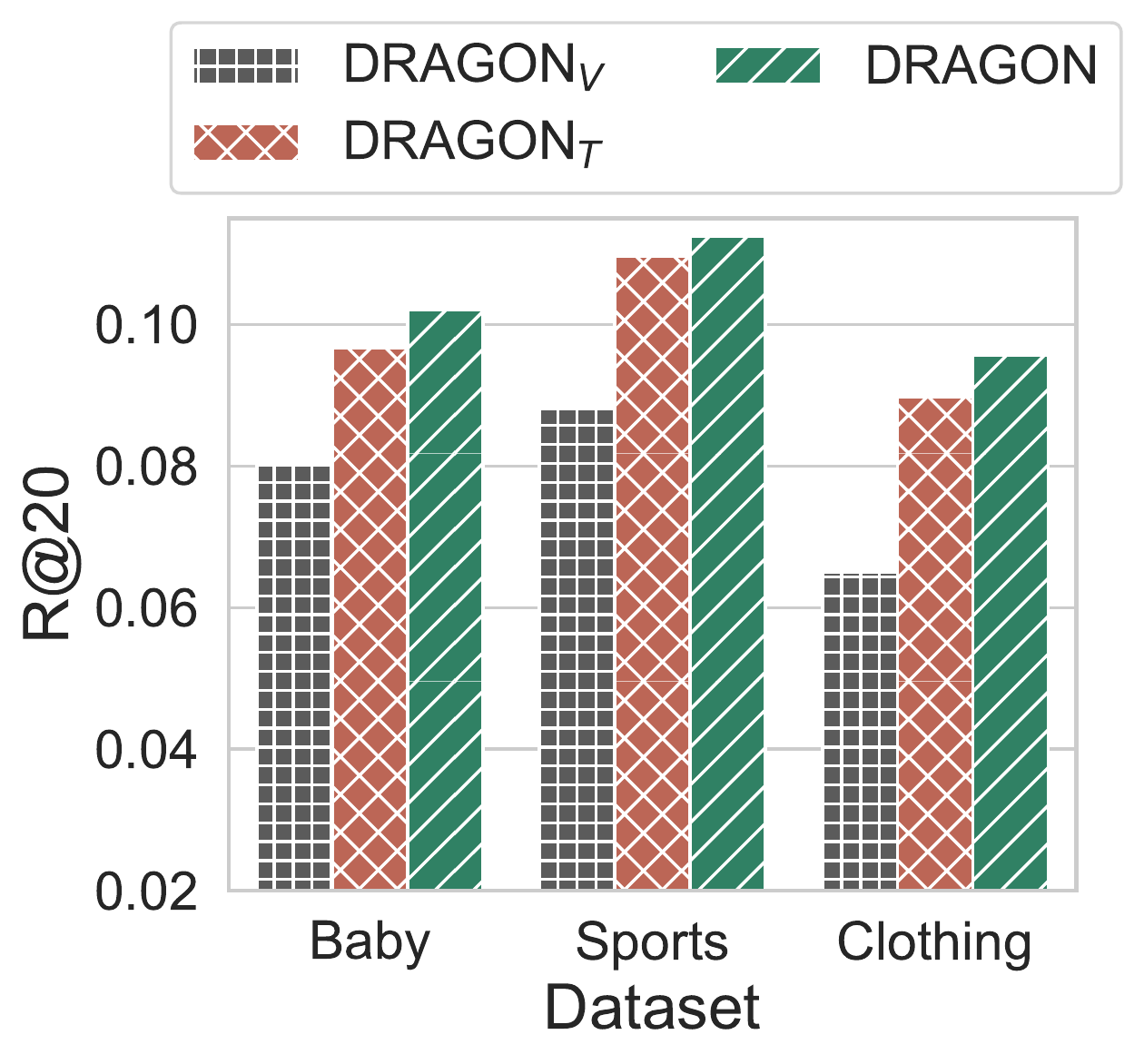}
\label{modality}
}
\caption{Ablation study of \mn.}
\end{figure}

The full version \mn adds item-item homogeneous graph to $\mn_{UU}$.
Fig.~\ref{components} shows our comparison results in terms of Recall@20. We have the following observations that show all components of \mn contribute to the performance: (1) LATTICE and SLMRec are the strongest baselines for multimodal recommendation, $\mn_{UI}$ excludes the dual representation learning and only utilizes our multimodal fusion can reach almost the same or even higher accuracy with the strong baselines. (2) The performance accuracy comparison result $\mn_{UI} < \mn_{UU} < \mn$ shows that adding user co-occurrence graph gets improvement compared with only utilizing the u-i graph. Based on $\mn_{UU}$, adding the item semantic graph could reach higher improvement. We observe that utilizing the user co-occurrence graph does not get much improvement for Baby. It suggests that the graph connectivity also influences the construction of the user co-occurrence graph. With more interactions between users and items, more co-occurrence patterns of user relations can be captured.

\subsubsection{Effect of different modality fusion method} 
We identify the modality fusion issue for previous works and utilize the direct concatenation with attention in our proposed model. We compare its performance with the fusion methods mentioned in \cite{wang2021dualgnn}\cite{zhang2021mining}. We replace the fusion of \mn with weighted sum~(denoted as $\mn_{sum}$), mean~(denoted as $\mn_{mean}$) and weighted max~(denoted as $\mn_{max}$) to demonstrate the superiority of concatenation fusion used in \mn.
Fig.~\ref{fusion} shows our comparison results in terms of Recall@20. 
Clearly, our fusion method outperforms other methods as it can capture the complementary information of each modality and attend information from all modalities for recommendation.

\subsubsection{Effect of single modality \textit{vs.} multi-modalities}
In Introduction, we reveal the performance of previous multimodal models might degrade under multimodal settings. 
Hence, we compare the performance of \mn under uni-modal and multimodal settings.
$\mn_{V}$, $\mn_{T}$ and \mn denote the models that utilize visual, textual, and both modalities respectively.
Fig.~\ref{modality} shows our comparison results in terms of Recall@20. We observe that: (1). Models with different single modality information have different performances. The textual modality performs better than the visual modality. 
(2). \mn with multimodal outperforms that utilizing single modality, it demonstrates that fusing different modalities of information can improve the performance of \mn. Thus, the fusion method in our proposed model is indispensable.

\begin{figure}[h]
  \centering
    \begin{subfigure}[Baby]{\label{baby}
    \includegraphics[width=0.9\textwidth]{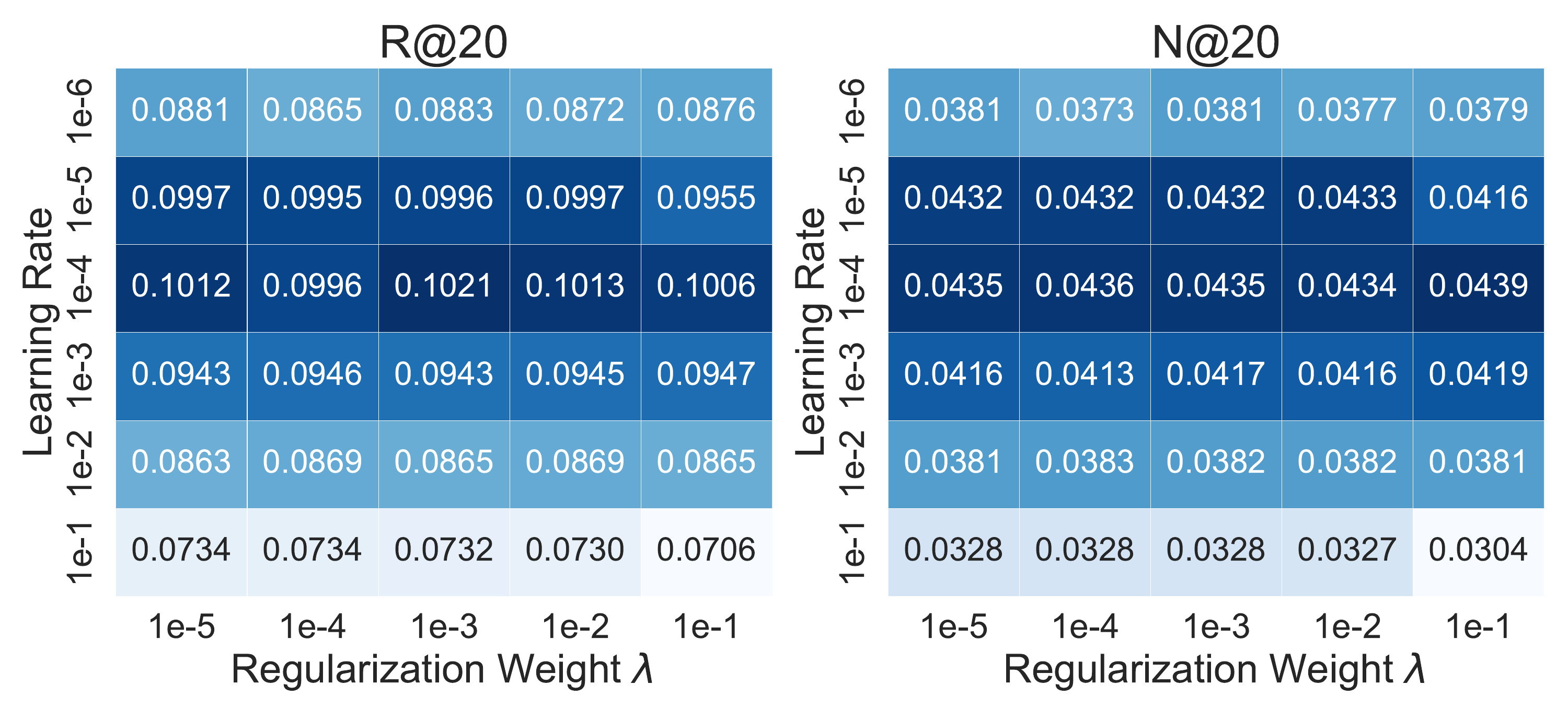}}%
 \end{subfigure}
    \begin{subfigure}[Clothing]{\label{clothing}
    \includegraphics[width=0.9\linewidth]{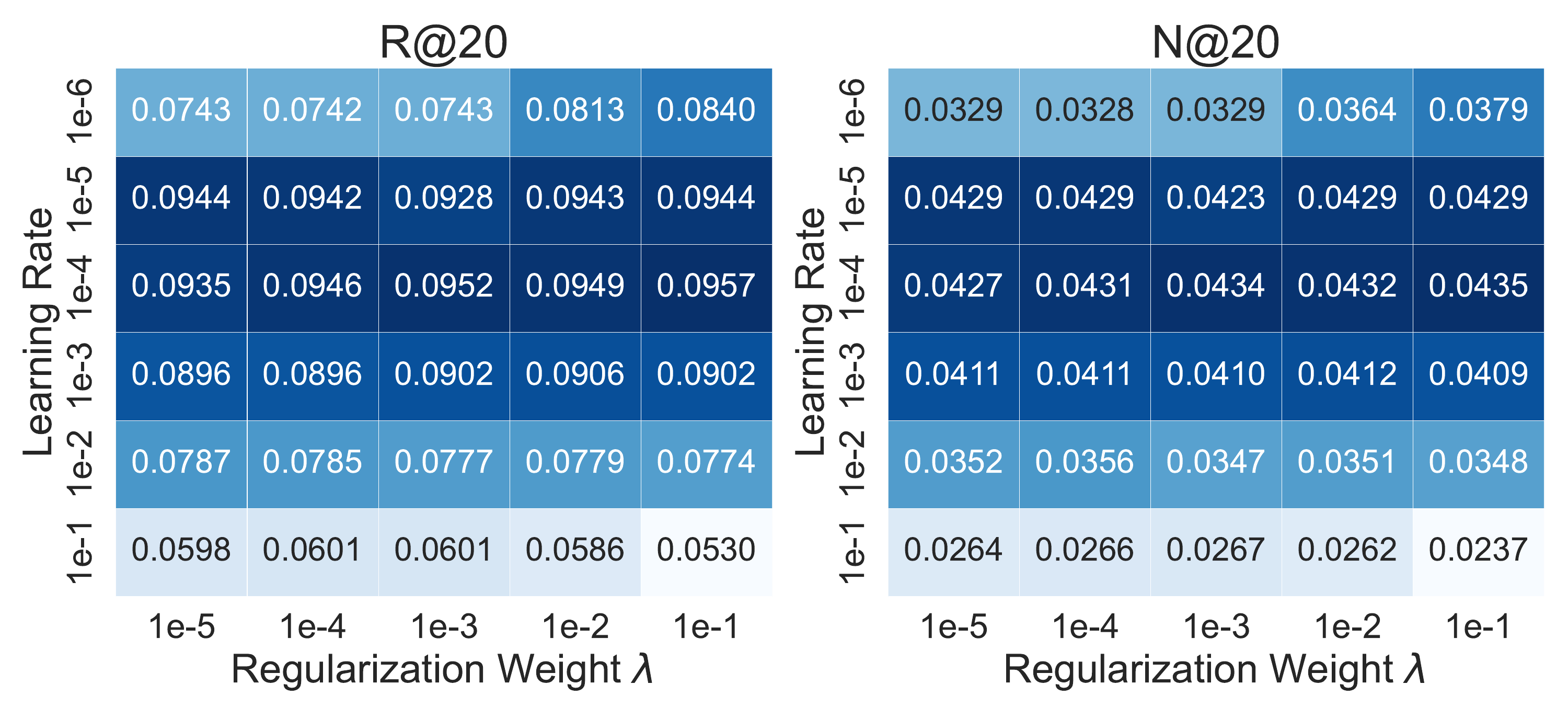}}%
    \end{subfigure}
    \caption{Sensitivity analyses on the \mn hyperparameters.}
\end{figure}
\subsection{Hyper-parameter Sensitivity Study}

To guide hyper-parameters selections of \mn, we perform the sensitivity study in terms of Recall@20 and NDCG@20. We vary the learning rate of \mn in \{1e-6, 1e-5, 1e-4, 1e-3, 1e-2, 1e-1\}, and vary the regularization weight $\lambda$ in \{1e-5, 1e-4, 1e-3, 1e-2, 1e-1\}. Fig.~\ref{baby} and Fig.\ref{clothing} show the performance of \mn under different combinations of the learning rate and regularization weight on Baby and Clothing datasets. 
From the figures, we have the following observations: (1). The suggested learning rate is 1e-4 which obtains the best result across all regularization weights. Except for the too-large and too-small learning rates, our model gets stable performance with the middle number of learning rates in \{1e-3, 1e-4, 1e-5\}. It further demonstrates the efficiency and stability of our model. (2). Compared with the learning rate, the performance is less sensitive to the regularization weight. However, searching the parameter sets of regularization weight in \mn also improves its performance. 

\section{Conclusion}
In this paper, we aim to solve the modality fusion issue and learn better representations for dyadic-related users and items. Therefore, we develop a novel model, named \mn, to learn the dual representations of users and items by constructing homogeneous and heterogeneous graphs. In particular, we first construct the modality-specific user-item bipartite graph to learn the modality features. After getting the representations of each modality, we utilize the late concatenation fusion method to learn the multimodal features. Then, we construct the user co-occurrence graph to capture the co-occurrence relations between users and the item semantic graph to capture the semantic relations between items. Therefore, we learn both inter- and intra-relations of the dyadic-related users and items. Finally, we conduct extensive experiments on three datasets to demonstrate the effectiveness of our proposed model. 
Our experimental finding on multimodal fusion could shed light on the design of future multimodal recommender systems.

%
%
%
%
%
\bibliographystyle{splncs04}
\bibliography{ref}
%




\end{document}